\documentclass[twocolumn]{IEEEtran}

\usepackage{amsmath,nicefrac}

\newcommand{\bin}[2]{\left(\begin{array}{@{}c@{}}#1\\#2\end{array}\right)}

\begin{document}
\title{Comment on ``New Results on Frame-Proof Codes and Traceability
  Schemes''} \author{Jan-\AA ke Larsson\thanks{J.-\AA.~Larsson is with
    the Department of Mathematics, Link\"oping University, SE-581 83
    Link\"oping, Sweden (e-mail: jalar@mai.liu.se).} and Jacob
  L\"ofvenberg\thanks{J.~L\"ofvenberg did this work while at the
    Department of Electrical Engineering, Link\"oping University,
    SE-581 83 Link\"oping, Sweden. J.~L\"ofvenberg is presently with
    the Department of Systems Development and IT-security, FOI:
    Swedish Defence Research Agency, SE-164 90 Stockholm, Sweden
    (e-mail: jaclof@foi.se).}}

  
\date{15 Aug 2008} \maketitle
  
\begin{abstract}
  In the paper ``New Results on Frame-Proof Codes and Traceability
  Schemes'' by Reihaneh Safavi-Naini and Yejing Wang [IEEE Trans.\
  Inform.\ Theory, vol.\ 47, no.\ 7, pp.\ 3029--3033, Nov.\ 2001],
  there are lower bounds for the maximal number of codewords in binary
  frame-proof codes and decoders in traceability schemes. There are
  also existence proofs using a construction of binary frame-proof
  codes and traceability schemes. Here it is found that the main
  results in the referenced paper do not hold.
\end{abstract}
  
\begin{keywords} fingerprinting, watermarking, frame-proof codes,
  traceability schemes
\end{keywords}
  
\section{Introduction}

We will examine the results in the paper "New Results on Frame-Proof
Codes and Traceability Schemes" by Reihaneh Safavi-Naini and Yejing
Wang \cite{rei01}. Frame-proof codes were introduced in \cite{bs95}
and is a technique to deter from illegal copying. The basic idea is
that somebody has a digital document they want to distribute to a
number of users, and to make it possible to trace illegal copies,
he/she incorporates small changes in the document. If a single user
makes a copy of his/her document it is simple to determine the
identity of the guilty user by examining the copy.

A stronger attack is if several users cooperate to create a new
document that is a combination of their copies, and it is here that
frame-proof codes are useful. Each user gets a copy of the digital
document that corresponds to a codeword in the frame-proof code. The
relation between the codeword and the document copy is that each
coordinate in the code decides what alternative is chosen in one of
the places where changes are allowed. It is further assumed that users
working together to create an unsolicited copy can in each changeable
position only choose among the alternatives given in their copies.
Described in terms of words in the code space, a group of users can
create any word which for every coordinate is equal to at least one of
the codewords belonging to them. The combinatorial properties of a
frame-proof code are such that, as long as the number of
redistributors is limited, they cannot create the codeword/document of
another user.
 
Traceability schemes were introduced in \cite{cfn94} and are in some
ways similar to frame-proof codes. A common scenario is a broadcast of
some digital data stream that is encrypted and available only to
authorized users. The stream is decrypted using a decoder containing
suitable decryption keys. In a traceability scheme there is a base set
$K$ of keys, of which each decoder contains a unique subset of size
$k$. A set of users may want to create a pirate decoder by using a
suitable combination of some of the keys in their decoders. In a
traceability scheme, if the number of users working together is
limited, any such created pirate decoder will be possible to trace to
at least one of the guilty users. The idea is that this property will
deter from creating pirate decoders.

In the discussed paper constant-weight codes are used to make bounds
and explicit constructions. These codes have length $l$, constant
weight $w$, minimum Hamming distance $2\delta$, and $c$ is the number
of cooperating, copy-distributing users. $H(x)$ denotes the binary
entropy function, and logarithms are in base 2. We will discuss
Theorems 6, 7, 10, and 11 from~\cite{rei01}.

\section{On Theorem 6}
\label{Th6}

Our initial concern is Theorem 6 on binary frame-proof codes. It
depends on the following displayed inequalities:
\begin{equation}
  \label{eq:1}
  \frac {\log l} l < \sigma \mbox{\ \ and\ \ } l
  >\left( 13+\sqrt{13^2+48\sigma} \right) /12\sigma.
  \tag{[1]:6}
\end{equation}
We quote Theorem 6 from \cite{rei01}:

{\em Theorem 6:}\ \ Let $q$ be a prime power. Suppose there exists a
$c$-frame-proof code with length $l \le q$, constant weight $w$, and
$c=l/w$. Then, for any $\sigma > 0$ and $l$ satisfying (\ref{eq:1}),
the maximum number of codewords $n$ satisfies
\begin{equation}
  \label{eq:2}
  n > \frac 1{q^{\delta - 1}} 2^{(H(\frac 1c)-\sigma)l}.
  \tag{[1]:13}
\end{equation}

There is no formal proof of this in \cite{rei01}, but we have studied
the discussion leading to Theorem 6 in some detail to reconstruct a
proof. We will not repeat the necessary steps here, but only mention
that the implication in Lemma 3 of \cite{rei01} is needed in the
reverse direction for the proof to go through. That this implication
is not an equivalence can be seen by using, for example, the code
$G=\{0011, 0110, 1100\}$ in Lemma 3, see \cite{rei01}.

In any case, the following counterexample shows explicitly that
Theorem 6 does not hold. We restate the upper bound from
\cite{ssw01}\footnote{The bound $n \le s^{\lceil \nicefrac lc \rceil
  }+2c-2$ mentioned in \cite{rei01} is not valid for $c$-frame-proof
  codes as the authors claim, see \cite{ssw01}. The bound \eqref{eq:3}
  used here is less restrictive but will still be contradicted by
  Theorem~6. We also note that the definitions of {\em feasible set}
  (called set of descendants in \cite{ssw01}) differ between
  \cite{rei01} and \cite{ssw01}, but it is easy to verify that in
  spite of this, the definitions of {\em frame-proof code} are
  equivalent.} on the number of codewords in a $c$-frame-proof code
over an alphabet of size $s$,
\begin{equation}
  \label{eq:3}
  n \le c\left(s^{\lceil \frac lc \rceil }-1\right).
\end{equation}

We will compare these bounds for code length $l=64$ and $c=2$. In
Theorem 6 we also need values for $q, \delta, w$ and $\sigma$. From
the relation $c=l/w$ we obtain $w=32$, and we will use $q=64$. Nothing
is said about $\delta$, but the lemmas leading to Theorem 6 require
that $\delta \ge 3$, so we will use $\delta=3$. The value for $\sigma$
must meet the inequalities in (\ref{eq:1}), and we use $\sigma =
7/64$.

To be in compliance with the prerequisites of Theorem 6 we must show
that there exists a 2-frame-proof code with these parameters. Let
$I$ be the unity matrix of size 3, and let $\overline 0$ and
$\overline 1$ be three-dimensional column vectors of zeros and ones
respectively. Let $\Gamma$ be a code with the code matrix
\begin{equation}
  \label{eq:4}
  \Gamma = I\rule{0pt}{2ex}^3 \overline 1\rule{0pt}{2ex}^{29}
  \overline 0\rule{0pt}{2ex}^{26},
\end{equation}
meaning the concatenation of three unity matrices, 29 one-vectors and
26 zero-vectors. This code has three codewords and is a 2-frame-proof
code.

Now everything is in place, and with the proposed values in Theorem 6
we get
\begin{equation}
  \label{eq:5}
  n > \frac 1{(2^6)^2} \times 2^{64 \times (1-\nicefrac{7}{64})}=2^{45}.
\end{equation}
Using the same $l$ and $c$ in (\ref{eq:3}) yields
\begin{equation}
 \label{eq:6}
  n \le 2(2^{32}-1)<2^{33},
\end{equation}
which clearly contradicts expression (\ref{eq:5}). Thus we conclude
that Theorem 6 in \cite{rei01} does not hold.

\section{On Theorem 7}
We now turn to the similar Theorem 7, which we quote from \cite{rei01}.  

{\em Theorem 7:}\ \ Let $q$ be a prime power. Suppose there exists a
$c$-traceability scheme with $l$ keys, $l \le q$, such that there are
$k$ keys in each decoder, and $c^2=2l/k$. Then, for any $\sigma > 0$
and $l$ satisfying (\ref{eq:1}), the maximum number of decoders $n$
satisfies
\begin{equation}
  \label{eq:7}
  n > \frac 1{q^{\delta - 1}} 2^{(H(\frac 1{c^2})-\sigma)l}.
  \tag{[1]:15}
\end{equation}

Again, there is no formal proof, but a similar exercise as for
Theorem~6 shows that Lemma~5 in \cite{rei01} is used in the reverse
direction. We proceed straight to the counterexample. We restate from
\cite{sw98} an upper bound on the number of decoders in a
$c$-traceability scheme, also given as expression (4) in \cite{rei01}.
In a $c$-traceability scheme it holds that
\begin{equation}
  \label{eq:8}
  n \le {\bin{l}{t}}\Big/{\bin{k-1}{t-1}},
  \tag{[1]:4}
\end{equation}
where $t=\lceil \frac kc \rceil$ and $k$ is the number of keys
contained in each decoder.

Let us choose $l=256$ and $c=4$. In Theorem 7 we also need values for
$q, \delta, k$ and $\sigma$. From the relation $c^2=\nicefrac{2l}k$ we
obtain $k=32$, and similarly to Section~\ref{Th6} we choose $q=256$,
$\sigma=\nicefrac9{256}$, and $\delta=3$.

Using $l=256$ and $k=32$ it is possible to construct a trivial
traceability scheme with eight decoders, each containing 32 keys and
no pair of decoders sharing any keys. This traceability scheme can
handle (at least) $c=4$ users working together to create a pirate
decoder.

Again everything is in place so we can use the proposed values in
expression (\ref{eq:7}), and this yields a lower bound on the
maximal number of decoders as
\begin{equation}
  \label{eq:10}
  n>2^{-16} \times 2^{256 \times (H(\nicefrac 1{16})-\nicefrac 9{256})}
  >2^{61}.
\end{equation}
The same $l$, $c$ and $k$ in expression (\ref{eq:8}) yields
\begin{equation}
  \label{eq:11}
  n \le {\bin{256}{8}}\Big/{\bin{31}{7}} < 2^{28},
\end{equation}
which clearly contradicts expression (\ref{eq:10}). Thus we conclude
that Theorem 7 in \cite{rei01} does not hold.

\section{On Theorems 10 and 11}
\label{sec:theorem-10-11}

Even if Theorems 6 and 7 do not hold, there is an explicit
construction underlying Theorems 10 and 11 in \cite{rei01}, also
providing lower bounds for the number of code words $n$. The claim is:

{\em Theorem 10:}\ \ For a given integer $c>1$, there exists a
$c$-frame-proof code for which the parameters are restricted by
\eqref{eq:1},
\begin{align}
  \label{eq:12}
  \sigma&=\tfrac12\left(H\left(\tfrac1c\right)-\tfrac1c\right),
  \tag{[1]:17}\\
  \label{eq:13} c&=\tfrac lw,\tag{[1]:18}\\
\intertext{and} 
  \label{eq:14}
  \log l&<\frac12\cdot\frac{c^2}{c-1}\sigma.
  \tag{[1]:20}
\end{align}

Unfortunately, there is no way to choose the parameters so that
\eqref{eq:12}, \eqref{eq:13} and~\eqref{eq:14} are simultaneously
satisfied. Furthermore, even if we fall back to the underlying
construction, we find ourselves in similar difficulties.

To see this, we start by inserting \eqref{eq:12} and \eqref{eq:13} in
(\ref{eq:14}), arriving at
\begin{equation}
  \label{eq:15}
  \log wc<\frac14\cdot\frac{c^2}{c-1}
  \left(H\left(\tfrac1c\right)-\tfrac1c\right).
\end{equation}
The inequality $\ln x\le x-1$ gives us
\begin{equation}
  \label{eq:25}
  H\left(\tfrac1c\right)\le\frac{\log{c}+\log{e}}c  
\end{equation}
which inserted in \eqref{eq:15} yields
\begin{equation}
  \label{eq:16}
  \log w+\log c<\frac14\cdot\frac{c^2}{c-1}
  \frac{\log c+ \log e-1}c.
\end{equation}
The required integer $c>1$ makes $\nicefrac{c}{(c-1)}\le2$ and
\begin{equation}
  \label{eq:17}
  \log w< \frac12(\log e-1-\log c)<0.
\end{equation}
That is, Theorem~10 enforces weight $w<1$. We can only conclude that
the theorem is invalid as it stands in \cite{rei01}.

We will now go into more detail in the proof of Theorem~10, to show
that also the underlying construction scheme fails. In place of
(\ref{eq:14}), this construction uses a more complicated expression:
the parameters must allow the existence of an integer $\delta>0$ such
that
\begin{equation}
  \label{eq:18} \left(1-\tfrac1c\right)w-1<\delta
  \le\left(H\left(\tfrac1c\right)-\tfrac1c-\sigma\right) \frac{l}{\log
    l}. \tag{[1]:19}
\end{equation} 
The left-hand inequality of~\eqref{eq:18} ensures that the code is a
frame-proof code, while the right-hand inequality ensures the desired
behavior of the number of codewords $n>2^{\nicefrac lc}$. The integer
$\delta$ is to be used in Theorem 8 and 9 of \cite{rei01} to show
\emph{existence} of a code with the desired properties. The inequality
(\ref{eq:14}) is claimed to guarantee existence of such a $\delta$,
but there is no motivation of this claim in \cite{rei01}. We will
perform a more thourogh examination here which will show that no
parameter values allow existence of such an integer $\delta>0$.

Inserting the conditions \eqref{eq:12} and \eqref{eq:13} in
\eqref{eq:18} we arrive at
\begin{equation}
  \label{eq:19} \left(1-\tfrac1c\right)w-1<\delta
  \le\tfrac12\left(H\left(\tfrac1c\right)-\tfrac1c\right)\frac{wc}{\log
    wc},
\end{equation} 
For the needed $\delta>0$ to exist, it is clear that the following
needs to be positive:
\begin{equation}
  \label{eq:20}
  f(w,c)=\tfrac12\left(H\left(\tfrac1c\right)-\tfrac1c\right)
  \frac{wc}{\log wc} -\left[\left(1-\tfrac1c\right)w-1\right],
\end{equation}
Using \eqref{eq:25}, we obtain
\begin{equation}
  \label{eq:21} f(w,c)\le\tfrac12(\log c+\log e-1)\frac w{\log
wc}-\left(1-\tfrac1c\right)w+1.
\end{equation}
This is positive if $w=1$. When $w\ge2>\nicefrac e2$, we have
$\log{w}>\log e-1$, so that \eqref{eq:21} simplifies to
\begin{equation}
  \label{eq:22} f(w,c)\le w\left(\dfrac1w+\dfrac1c-\dfrac12\right).
\end{equation}
Clearly, $0<f(w,c)$ needs either $w=1$ or $w=2$ or $c=2$, or that
$(c,w)$ is one of $(3,3)$, $(4,3)$, $(5,3)$, $(3,4)$ or $(3,5)$. We
analyze these four cases separately:
\begin{enumerate}
\renewcommand{\theenumi}{\alph{enumi}}
\renewcommand{\labelenumi}{\theenumi)}
\item When $w=1$, using \eqref{eq:25}, the right-hand inequality of
  \eqref{eq:19} simplifies to
  \begin{equation}
    \label{eq:9}
    \delta
    \le\tfrac12\left(1+\frac{\log e -1}{\log c}\right)< 1.
  \end{equation}
  Consequently no such integer $\delta>0$ exists.
\item When $w=2$, the expression \eqref{eq:21} is
\begin{equation}
  \label{eq:24} f(2,c)\le\frac{\log c+\log e-1}{\log c+1}-1+\frac2c
  =\frac{\log e-2}{\log c+1}+\frac2c,
\end{equation}
which is positive for $c=2$, decreases to a negative minimum and then
increases to 0 as $c$ tends to infinity; it is positive 
only for $c<19$, and for these values the left-hand side of
\eqref{eq:19} is also less than 1.
\item When $c=2$ we have $H(\nicefrac12)=1$, so that
\begin{equation}
  \label{eq:23} f(w,2)=\frac{w}{2\log2w}-\frac w2+1.
\end{equation}
This is strictly decreasing, and positive only for $w=2$ and $w=3$;
for these values the right-hand side of \eqref{eq:19} is also less
than 1.
\item The remaining candidates $(w,c)=(3,3)$, $(4,3)$, $(5,3)$,
  $(3,4)$, and $(3,5)$ all give $f(w,c)$ a negative value.
\end{enumerate}

That is, there is no combination of $w$ and $c>1$ that allows an
integer $\delta>0$ obeying expression~\eqref{eq:18}. A little more
effort will show that for some combinations of $w$ and $c$, there are
$\delta>0$ that obey \emph{either} the left or the right inequality,
but not both. For these values of $\delta$, the constructed codes are
\emph{either} guaranteed to be $c$-frame-proof codes \emph{or}
guaranteed to have a number of code words $n>2^{\nicefrac lc}$, but
the constructed codes are \emph{never} guaranteed to have \emph{both}
properties.

Similar reasoning holds for Theorem 11, with the sole difference that
the parameter $c$ is inserted squared in the equivalent of expressions
(\ref{eq:14}) and (\ref{eq:18}). And in a similar fashion, the
construction does not allow construction of a traceability scheme that
is guaranteed to have the desired number of decoders.

\section{Conclusions}
\label{sec:conclusions}
We have found that Theorems 6 and 7 of \cite{rei01} for some choices
of parameter values violate previously published upper bounds. They
can clearly not be valid as they stand.

We have also found that Theorems 10 and 11 of \cite{rei01} are
invalid. There are no parameter values that fulfill the given bounds
simultaneously; the requirements are simply too restrictive. In other
words, the theorems cannot be used to prove existence of codes with
the desired properties. The underlying construction gives codes which
may have one of the two desired properties, but the codes are never
guaranteed to have both.

\bibliographystyle{IEEEtran} 
\bibliography{ieee}

\end{document}